\documentclass{an}
\usepackage{graphicx}
\usepackage{times}
\usepackage{fancyhdr}
\begin{document}

\title{Effects of cosmological magnetic helicity on the CMB}

\author{Tina Kahniashvili \inst{1,2}}
\institute{Department of Physics, Kansas State University, Manhattan, KS, USA
\and
Center for Plasma Astrophysics, Abastumani Astrophysical Observatory, Tbilisi, Georgia}

\date{Received $<$date$>$;
accepted $<$date$>$;
published online $<$date$>$}

\abstract{In this talk I present a short review of primordial
magnetic helicity effects on Cosmic Microwave Background (CMB)
temperature and polarization anisotropies. These effects
allow us to test for cosmological magnetic helicity, however, very accurate
 CMB fluctuation data  is required. This scheme for magnetic helicity
detection is valid only for
 a cosmological magnetic field with a present  amplitude larger than
 $10^{-9}-10^{-10}$ Gauss.
}

\correspondence{tinatin@phys.ksu.edu}

\maketitle

\section{Introduction}

Recent astrophysical observations indicate that the
magnetic fields in the Sun and some galaxies and  clusters of galaxies
 might have an helical structure (for reviews see (Widrow 2002;
Vall{\'e}e 2004)). One promising possibility to explain the
observed magnetic helicity is assuming the presence of {\it seed}
helical magnetic fields. Several mechanisms have been proposed
for generating   such a primordial helical magnetic field during
an  early epoch of the universe
 (Cornwall 1997; Giovannini \& Shaposhnikov 1998; Giovannini 2000;
Field \& Carroll 2000; Vachaspati 2001; Sigl 2002; Semikoz \&
Sokoloff 2005; Campanelli \& Giannotti 2005). Magnetic helicity
can be also be generated via helical turbulent motions if there
is an inverse cascade (Brandenburg 2001; Chistensson, Hindmarsh \&
Brandenburg 2005; Verma \& Ayyer 2003; Boldyrev \& Cattaneo 2004).
On the other hand, primordial helicity influences
magneto-hydrodynamical processes in the early plasma as well as
cosmological perturbation dynamics (Vichniac \& Cho 2001;
Brandenburg 2001; Kleorin et al. 2003; Subramanian 2002; Vishniac,
Lazarian \& Cho 2003; Subramanian \& Brandenburg 2004; Banerjee \&
Jedamzik 2004; Subramanian \& Shukurov 2005).

To preserve large-scale  isotropy, a  seed magnetic field (and
so  magnetic helicity) has to be
small enough to allow  treating the average energy density and mean
helicity of the magnetic field  as first order in perturbation theory.

Even if the energy density of a primordial magnetic field
$B^2/(8\pi)$ is five or six  magnitude
less than that of radiation (CMB photons), taking into account that
CMB anisotropy measurements have the same order
of accuracy, $10^{-6}-10^{-5}$, we  expect that such a cosmological magnetic
 field
would leave detectable traces in CMB temperature or polarization
anisotropies, see (Lewis 2004; Giovannini 2005) and references
therein.

Here I focus on the effects on CMB temperature and polarization
anisotropies induced by magnetic helicity. This talk is based on
results obtained in collaboration with C. Caprini, R. Durrer, G.
Gogoberidze, A. Kosowsky, G. Lavrelashvili, A. Mack, and B.
Ratra.  (Mack, Kahniashvili \& Kosowsky 2002; Caprini, Durrer \&
Kahniashvili 2004; Kosowsky et al. 2005; Kahniashvili \& Ratra
2005; Kahniashvili, Gogoberidze \& Ratra 2005). We find that
magnetic helicity generates vector and tensor metric
perturbations and as a result affects all CMB fluctuations. In
particular:  (i) magnetic helicity reduces the amplitudes of
parity-even CMB fluctuation power spectra in comparison to the
case of a non-helical magnetic field (Caprini et al. 2004;
Kahniashvili \& Ratra 2005); (ii) the Faraday rotation of the CMB
polarization plane is strongly dependent on the average energy
density of the magnetic field and  is independent of magnetic
helicity, see Kosowsky et al. (2005); (iii) magnetic helicity
 induces  parity-odd cross-correl
 ations of the CMB fluctuations,
which vanish for the case of a magnetic field without helicity
(Pogosian, Vachaspati \& Winitski 2002; Caprini et al. 2004;
Kahniashvili \& Ratra 2005) \footnote{This is not
  true for a homogeneous
magnetic field, Scoccola, Harrari \& Mollerach (2004)}; and (iv)
magnetic helicity generates  circularly polarized stochastic
gravitational waves, Kahniashvili et al. (2005).

\section{Magnetic source for metric perturbations}

We assume the existence of a cosmological magnetic field generated during or
prior to the radiation-dominated epoch, with the energy density of the
 field a first-order perturbation to the standard
Friedmann-Lema\^\i tre-Robertson-Walker
homogeneous cosmological spacetime model.
Neglecting fluid back-reaction onto the magnetic field,
the spatial and temporal dependence of the
field separates,
${\mathbf B}(t,{\mathbf x})={\mathbf B}({\mathbf x})/a^2$; here $a$ is the
 cosmological scale factor.
As a phenomenological normalization of the magnetic field,
we smooth the field on a comoving length $\lambda$ with a
Gaussian smoothing kernel $\propto \mbox{exp}[-x^2/\lambda^2]$
 to obtain the smoothed
magnetic field
with average value of squared magnetic field
${B_\lambda}^2 \equiv \langle {\mathbf B}({\mathbf x})
\cdot {\mathbf B}({\mathbf x})\rangle |_\lambda$ and
magnetic helicity
${H_\lambda}^2 \equiv \lambda
| \langle{\mathbf B}({\mathbf x}) \cdot
[{\mathbf \nabla} \times {\mathbf B} ({\mathbf x})] \rangle|_\lambda$.

We also assume that the primordial plasma is a perfect conductor
on all scales larger than the Silk damping wavelength $\lambda_S$
(the thickness of the last scattering surface) set by photon and
neutrino diffusion. We model magnetic field  damping by an
ultraviolet cut-off wavenumber $k_D=2\pi/\lambda_D$, Subramanian
\& Barrow (1998),
\begin{eqnarray}
&&\left({k_D \over {\rm Mpc}^{-1}}\right)^{n_B + 5}
\approx\nonumber\\
&& \approx 2.9\times 10^4
  \left({B_\lambda\over 10^{-9}\,{\rm G}}\right)^{-2}
  \left({k_\lambda\over {\rm Mpc}^{-1}}\right)^{n_B + 3} h.
\label{kd}
\end{eqnarray}
Here $n_B$ is the spectral index of the symmetric part of the
 magnetic field power spectrum (see Eq.~(\ref{energy-spectrum-H}) below),
 $h$ is the Hubble constant in units of
$100$~km sec${}^{-1}$ Mpc ${}^{-1}$, $k_\lambda = 2\pi/\lambda$ is
the smoothing wavenumber,
 and $\lambda_D \ll \lambda_S$.
This assumes that magnetic field damping is due to the damping of
Alfv\'en waves
from photon viscosity.

Assuming that the stochastic magnetic field is Gaussianly
distributed, and accounting for the possible helicity of the
field, the magnetic field spectrum in wavenumber space is,
Pogosian et al. (2002),
\begin{eqnarray}
&&\langle B^\star_i({\mathbf k})B_j({\mathbf k'})\rangle
=\nonumber\\
&&=(2\pi)^3 \delta^{(3)}
({\mathbf k}-{\mathbf k'}) [P_{ij}({\mathbf{\hat k}}) P_B(k)  +
i \epsilon_{ijl} \hat{k}_l P_H(k)].
\label{spectrum}
\end{eqnarray}
Here $P_{ij}({\mathbf{\hat k}})\equiv\delta_{ij}-\hat{k}_i\hat{k}_j$
is the transverse plane projector with unit wavenumber components
$\hat{k}_i=k_i/k$, $\epsilon_{ijl}$ is the antisymmetric symbol, and
 $\delta^{(3)}({\mathbf k}-{\mathbf k'})$ is the Dirac delta function.
$P_B(k)$ and $P_H(k)$ are the symmetric and
helical parts of the magnetic field power
spectrum, assumed to be simple power laws on large scales,
\begin{eqnarray}
P_B(k) &\equiv & P_{B0}k^{n_B}=
\frac{2\pi^2 \lambda^3 B^2_\lambda}{\Gamma(n_B/2+3/2)}
(\lambda k)^{n_B},\nonumber\\
P_H(k) &\equiv &P_{H0}k^{n_H}=
\frac{ 2\pi^2 \lambda^3 H^2_\lambda}{\Gamma(n_H/2+2)}
(\lambda k)^{n_H},
\label{energy-spectrum-H}
\end{eqnarray}
and vanishing on small scales when $k>k_D$. Here $\Gamma$ is the
Euler Gamma function. These power spectra are generically
constrained by $P_B(k)\geq |P_H(k)|$, Caprini et al. (2004), which
implies $n_H> n_B$.
 In addition, finiteness of the magnetic
field energy density requires $n_B > -3$ (to prevent an infrared
divergence of magnetic field energy density).  Finiteness of the magnetic
field average helicity requires $n_H > -4$; this is  automatically
satisfied as a consequence of $n_H> n_B >-3$.

To obtain the magnetic field source  terms in the equations for vector
(transverse
peculiar velocity)  and tensor (gravitational waves) metric perturbations
  we need to
extract the transverse vector and tensor parts of the magnetic
field stress-energy tensor $\tau_{ij}({\mathbf{k}})$. This is done
through $ \Pi_{ij}^{(V)}({\mathbf{k}})=(P_{ib}({\mathbf{\hat k}})
\hat{k}_j+P_{jb}({\mathbf{\hat k}})\hat{k}_i)\hat{k}_a
\tau_{ab}({\mathbf{k}}) $ (for vector perturbations) and  $
\Pi_{ij}^{(T)} ({\mathbf {k}})=[ P_{ia}({\mathbf{\hat
k}})P_{jb}({\mathbf{\hat k}}) -\frac{1}{2} P_{ij}({\mathbf{\hat
k}}) P_{ab}({\mathbf{\hat k}}) ]\tau_{ab}({\mathbf{ k}})$ (for
tensor perturbations); for details see Mack et al. (2002).

 For vector perturbations the $\Pi_{ij}^{(V)}$ tensor is related to the
vector (divergenceless and transverse) part of the Lorentz force
$L_i^{(V)}({\mathbf{k}}) = k_j \Pi_{ij}({\mathbf{k}})
 =P_{ib}({\mathbf{\hat k}}) {k}_a
\tau_{ab}({\mathbf{k}}) $. For the normalized Lorentz force
vector,  $\Pi_i \equiv L_i^{(V)}/k$, the general spectrum of
$\langle\Pi^\star_i({\mathbf k})\Pi_j ({\mathbf k'})\rangle $ in
wavenumber space is similar to Eq.~(\ref{spectrum}), and has two
parts, symmetric  and helical. Both contain  contributions from
$P_H(K)$, and so magnetic helicity affects the vector magnetic
source term, Kahniashvili \& Ratra (2005).

The tensor magnetic source term  is obtained  through the
 source two-point function
$\langle \Pi_{ij}^{(T)\star} ({\mathbf k}) \Pi_{lm}^{(T)}
({\mathbf k^\prime})\rangle $.  This is determined by the magnetic
field two-point  function, and like the vector magnetic source
term has symmetric and helical parts (Caprini et al. 2004;
Kahniashvili et al. 2005). The  helical (parity-odd) piece
results in circular polarization of the induced gravitational
waves. For a maximally helical magnetic field with $P_H(k) \simeq
P_B(k)$, the polarization degree is high enough to allow us to
consider the  possibility of  testing for magnetic helicity
through a measurement  of the polarization of relic gravitational
waves (this might be possible with future gravitational waves
detectors); see Kahniashvili et al. (2005).

In both cases (vector and tensor perturbations)
the contribution of magnetic field
helicity to
the symmetric part of the magnetic source is negative.
It is clear that the magnetic source terms
vanish on scales smaller than the cutoff scale
 $\lambda_D$ because of magnetic field damping.
Depending on spectral
indexes $n_B$ and $n_H$ the magnetic source terms are dominated either by
small wavenumber, $\propto k^{2n_B+3}$ for $n_B<-3/2$, or by the high frequency cut-off and so  $\propto k_D^{2n_B+3}$
for $n_B>-3/2$.

\section{CMB anisotropies}

For our computations we use the formalism given in  Mack et al.
(2002), extending it to account for magnetic field helicity. To
compute CMB temperature and polarization anisotropy power spectra
we use the total angular momentum method of Hu \& White (1997).
Our analytical approximations are  deriven  in (Caprini et al.
2004; Kahniashvili \& Ratra 2005).

\subsection{Parity-even CMB fluctuations}
Cosmological magnetic field vector and tensor mode contributions
to CMB fluctuations at large angular scales (with multipole number
$l<100$)
 are of the same order of magnitude, Mack et al. (2002),
while for small angular scales  (where $l>100$)  CMB fluctuations are
dominated by the vector mode contribution because of gravitational wave
damping (Lewis 2004).

The  complete parity-even CMB fluctuation power spectra
may be expressed as,
 \begin{equation}
C^{{\mathcal X}{\mathcal X}^\prime}_l=
C^{{\mathcal X}{\mathcal X}^\prime}_{(S)l}-
C^{{\mathcal X}{\mathcal X}^\prime}_{(A)l}, 
\label{decomposition_Cl} \end{equation}
where  $\mathcal X$ is either $\Theta$, $E$, or  $B$,
which represent respectively the temperature, $E$-polarization, and
$B$-polarization anisotropies, and $C^{{\mathcal X}{\mathcal
X}^\prime}_{(A)l}$ are the antisymmetric power spectra induced by magnetic
helicity. The minus sign reflects
the negative contribution of
 magnetic helicity  to the total parity-even CMB
fluctuation  power spectra (from terms proportional to
$ \int d^3\!p\,P_H(p)P_H(|{\bf k- p}|)$).
For large angular scales this result
holds for both (vector and  tensor) modes, while for $l>100$ it applies only for vector perturbations.
The fractional differences $\kappa^{{\mathcal X}{\mathcal X'}}_l \equiv
1-C_{(A)l}^{{\mathcal X}{\mathcal X}^\prime}/
C_{(S)l}^{{\mathcal X}{\mathcal X}^\prime}$, where
$0<\kappa^{{\mathcal X}{\mathcal X'}}_l<1$, can be used
to characterize the
 reduction of the parity-even CMB fluctuation power
spectra amplitudes as a consequence of non-zero magnetic helicity. The ratio
$C_{(A)l}^{{\mathcal X} {\mathcal X}^\prime}/ C_{(S)l}^{{\mathcal X}
{\mathcal X}^\prime}$ may be expressed  in
terms of $P_{0H}/P_{0B}$ and spectral indexes $n_H$ and $n_B$
 (Caprini et al. 2004; Kahniashvili \& Ratra 2005).
While the reduction of parity-even power spectra amplitudes are
significant for a maximally helical field, the most interesting effects
from magnetic helicity are the generation of parity-odd CMB fluctuations,
such as the cross-correlations between temperature and $B$-polarization, and
 between $E$- and $B$-polarizations.

\subsection{Parity-odd CMB fluctuations}
Magnetic helicity induces  parity-odd cross correlations
between the $E$- and  $B$-polarization anisotropies,
 as well as between temperature
and $B$-polarization anisotropies, (Pogosian et al. 2002; Caprini
et al. 2004; Kahniashvili \& Ratra 2005). Such off-diagonal
parity-odd cross correlations occur also in the case of an
homogeneous magnetic field from the  Faraday rotation effect,
Scoccola et al. (2004), but not in the case of Faraday rotation in
a stochastic magnetic field, even one with non-zero helicity,
Kosowsky et al. (2005).
 Faraday rotation measurements used to measure  a magnetic field
amplitude  cannot be used to detect magnetic helicity (Ensslin \&
Vogt 2003; Campanelli et al. 2004; Kosowsky et al. 2005).
 A possible way of detecting
magnetic helicity directly from CMB fluctuation data
 is  to detect the
parity-odd CMB fluctuation cross-correlations or/and
to detect the effects magnetic  helicity has on
 parity-even CMB fluctuations, Kahniashvili \& Ratra (2005).

\subsubsection{Temperature-$B$-polarization cross-correlations}
At large angular scales ($l<100$) where the contribution from
the tensor mode is significant, for $n_B+n_H>-2$ the  vector mode
${C_l^{\Theta B (V)}}$ and the tensor mode
${C_l^{\Theta B (T)}}$
have the same $l$ dependence $\propto l^2$.
For all other values of
spectral indexes $n_{B}$ and $n_{H}$, the growth rate (with $l$) of
 ${C_l^{\Theta B (V)}}$  is faster than
${C_l^{\Theta B (T)}}$. The ratio between  temperature--$B$-polarization
signals from vector and tensor modes is independent
of the amplitudes
of the average magnetic field ($B_\lambda$) and
average magnetic helicity ($H_\lambda$).

For small angular scales ($l> 100$) where the tensor mode signal vanishes, for
a maximally helical magnetic field with $n_H \simeq n_B$, due
to the suppression factor $L_{\gamma,{\rm{dec}}}/\eta_0$ (where
$L_{\gamma,{\rm{dec}}}$ is the photon mean free path  at decoupling and
 $\eta_0$ conformal time today)
the temperature-$E$-polarization cross-correlation power spectrum,
$C_l^{\Theta E}$, is smaller than the  temperature-$B$-polarization
cross-correlation power spectrum,
$C_l^{\Theta B}$, but both are $ \propto l^2$, if $n_B+n_H>-5$. The same
suppression factor makes $C_l^{\Theta B}$   smaller than
$C_l^{\Theta \Theta}$. For an arbitrary helical field
 $C_l^{\Theta B}/C_l^{\Theta E}$  depends on the ratio $(P_{H0}/
P_{B0})k_D^{n_H-n_B}$ and order unity  prefactors that depend on
$n_B$ and  $n_H$.
 A dependence on $l$ appears
only if $n_B+n_H<-5$, when  the ratio, $C_l^{\Theta
B}/C_l^{\Theta E}$ decreases as  $\propto l^{n_B+n_H+5}$,
Kahniashvili \& Ratra (2005).

\subsubsection{$E$ and $B$-polarization cross-correlations}
For a tensor mode signal at large angular scales ($l<100$),
 $E$- and $B$-polarization cross-correlation $C_l^{EB}$
is of the same order of magnitude as the tensor
mode temperature--$B$-polarization anisotropy cross-correlation
 spectrum, $C_l^{\Theta B}$, Caprini et al. (2004).
 The situation is different for a vector mode which survives up to small
angular scales (e.g., Subramanian \& Barrow 1998; Mack et al.
2002; Lewis 2004; Giovannini 2005).
  In this case,
the $E$- and $B$-polarization anisotropy cross-correlation power
spectrum has a suppression factor of $kL_{\gamma, \rm{dec}}$
implying that $C_l^{EB} \ll C_l^{\Theta B}$. This is
consistent   with the result of Hu and White (1997).

\section{Conclusion}

I have discussed  how cosmological magnetic helicity affects
 CMB fluctuations.
 Even for a cosmological magnetic field with maximal helicity
 such effects may be detectable  only  if the current magnetic
 field amplitude is at least
$10^{-10}$ or $10^{-9}$ G on Mpc scales. A non-helical seed field
of this amplitude can be generated during inflation (Ratra 1992;
Bamba \& Yokoyama 2004). A cosmological magnetic field generates
a $B$-polarization signal via induced vector and/or tensor modes,
so   detection of such a signal may indicate the presence of a
cosmological magnetic field.
 However, it has to
be emphasized that a $B$-polarization anisotropy signal can also
  arise in other ways, such as from primordial tensor perturbations,
gravitational lensing, or Faraday rotation of the CMB anisotropy
polarization plane; for a review see  Subramanian (2004).
 The peak position of $B$-polarization anisotropy power spectrum,
$l^2 C_l^{BB}$,  may help to identify the
$B$-polarization source. For example, cosmological-magnetic-field-induced
 tensor perturbations only contribute on large
angular scales $l<100$, while $B$-polarization anisotropy from
gravitational lensing has a peak  amplitude   $l^2 C_l^{BB} \sim
10^{-14}$ at $l\sim 1000$ Challinor \& Lewis (2005). The Faraday
rotation $B$-polarization anisotropy signal from a field with
$B_\lambda =10^{-9}$ G (at $\lambda=1$ Mpc)
 and spectral index $n_B=-2$ peaks at a substantially smaller
 scale $l \sim 10^4$
with a frequency-dependent peak amplitude  $l^2 C_l^{BB} \sim
10^{-12}$ (at $10$ GHz) and  $l^2 C_l^{BB} \sim 10^{-14}$ (at $30$
GHz); see Fig.1. A non-helical cosmological
magnetic field with $B_\lambda = 10^{-9}$ G at $\lambda=1$ Mpc
induces a $B$-polarization anisotropy signal via the vector
perturbation mode with a peak amplitude $l^2 C_l^{BB} \sim
10^{-13}$
  at $l\sim 1000$, Lewis (2004).
 We have shown that a
magnetic field with maximal helicity results in the reduction
 of the $B$-polarization anisotropy
 signal on all scales by a factor of $1/3$ for  $-3/2 < n_B \simeq n_H$,
relative to the  non-helical magnetic field case, Kahniashvili \&
Ratra (2005).
\begin{figure}
\resizebox{\hsize}{!}
{\includegraphics[angle=270,width=2.5in]{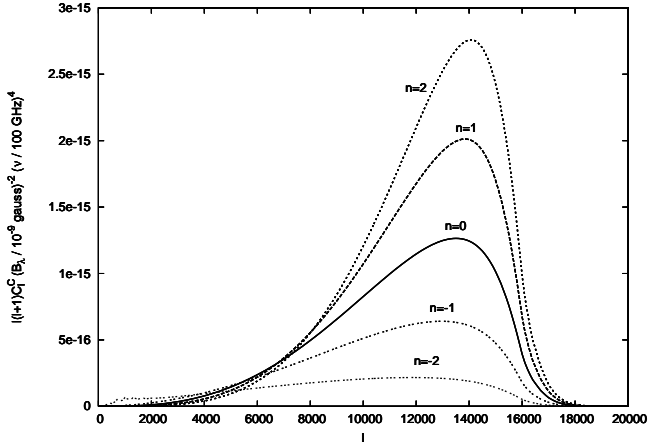}} \caption{The
$C$-polarization power spectrum of the microwave background
induced by the Faraday rotation effect, Kosowsky et al. (2005).
The curves in order of decreasing amplitude on the right side of
the plot correspond to magnetic field power spectral indices
$n_B=2$, 1, 0, $-1$, and $-2$. The magnetic fields have been
normalized to a nanogauss at the smoothing scale $\lambda = 1$
Mpc.} \label{label1}
\end{figure}
Summarizing we argue that
cosmological magnetic helicity in the case of a magnetic field larger
than $10^{-9}$ G  affects CMB anisotropies,
 in addition to the  effects
 it has on MHD dynamo amplification and processes in the early universe
(Cornwall 1997; Banerjee \& Jedamzik 2004). To measure primordial
magnetic helicity through these effects very accurate CMB
fluctuation data is required.

\acknowledgements
The author thanks the organizers of the conference
{\it The Origin and Evolution of Cosmic Magnetism} for hospitality, and the
SKA project for  partial support to attend the conference.
The author acknowledges her collaborators C. Caprini, R. Durrer,
G. Gogoberidze, A. Kosowsky, G. Lavrelashvili, A. Mack, and B. Ratra,
and thanks A.~Brandenburg, A. Dolgov, M. Giovannini, D. Grasso,
K. Jedamzik, and T. Vachaspati for discussions.
 This work is supported by DOE
EPSCoR grant DE-FG02-00ER45824.


\begin{thebibliography}{}

\bibitem{} Bamba, K., Yokoyama, J.:
2004, \ Phys. \ Rev. \ D {\bf 69}, 043507

\bibitem{}Banerjee, R., Jedamzik, K.: 2004, Phys.\ Rev.\ D {\bf 70},
123003

\bibitem{}Boldyrev, S., Cattaneo, F.: 2004, Phys. Rev. Lett. {\bf 92},
144501

\bibitem{}Brandenburg, A.: 2001, Astrophys.~J. {\bf 550}, 824

\bibitem{} Campanelli, L., Dolgov, A.D., Giannotti, M., Villante, F.L.: 2004, \ Astrophys.  \ J. {\bf 616}, 1

\bibitem{} Campanelli, L. and Giannotti, M.:  2005, astro-ph/0508653

\bibitem{} Caprini, C., Durrer, R., Kahniashvili, T.: 2004,
Phys.\ Rev.\ D {\bf 69}, 063006

\bibitem{}Challinor, A., Lewis, A.: 2005,  astro-ph/0502425

\bibitem{}Christensson, M., Hindmarsh, M., Brandenburg, A.:
 2005, Astron. Nachr. {\bf 326}, 393

\bibitem{}Cornwall, J.M.: 1997, Phys. \ Rev. \ D {\bf 56},
6146

\bibitem{}Ensslin, T., Vogt, C.: 2003, Astron. \& Astrophys. {\bf 401},
835

\bibitem{}Field, G.B., Carroll, S.M.: 2000, Phys. Rev. D {\bf 62}, 103008

\bibitem{}Giovannini, M.: 2000, Phys. \ Rev. \ D. \ {\bf 61}, 063004

\bibitem{} Giovannini, M.: 2005, astro-ph/0508544

\bibitem{}Giovannini, M., Shaposhnikov, M.: 1998, Phys. Rev. D {\bf 57}, 2186

\bibitem{} Hu, W., White, M.: 1997,
Phys. Rev. D {\bf 56}, 596

\bibitem{} Kahniashvili, T., Gogoberidze, G., Ratra, B.: 2005,
Phys. Rev. Lett. {\bf 95}, 151301

\bibitem{} Kahniashvili, T., Ratra, B.: 2005, Phys. Rev. D {\bf 71}, 103006

\bibitem{} Kleeorin, N., Moss, D., Rogachevskii, I., Sokoloff,
D.:  2003, Astron. \& Astrophys. {\bf 400}, 9

\bibitem{} Kosowsky, A., Kahniashvili, T., Lavrelashvili, G., Ratra, B.: 2005, Phys. Rev. D {\bf 71}, 043006

\bibitem{} Lewis, A.: 2004, Phys. Rev. D {\bf 70}, 043011

\bibitem{}Mack, A., Kahniashvili, T., Kosowsky, A.: 2002,
Phys.\ Rev.\ D {\bf 65}, 123004

\bibitem{}Pogosian, L., Vachaspati, T., Winitzki, S.: 2002,
Phys. \ Rev. \ D {\bf 65}, 083502

\bibitem{} Ratra, B.: 1992, Astrophys. J. Lett. {\bf 391}, L1

\bibitem{}Scoccola, C., Harari, D., Mollerach, S.: 2004,
\ Phys. \ Rev. \ D {\bf 70}, 063003

\bibitem{}Semikoz, V., Sokoloff, D.: 2005,
Astron. \& Astrophys. {\bf 433}, L53

\bibitem{}Sigl, G.: 2002, Phys. Rev. D {\bf 66}, 123002


\bibitem{} Subramanian, K.: 2002, astro-ph/0204450

\bibitem{} Subramanian, K.: 2004, astro-ph/0411049

\bibitem{} Subramanian, K., Barrow, J.: 1998, Phys. Rev. D {\bf 58}, 083502

\bibitem{}Subramanian, K. Brandenburg, A.: 2004,
Phys. \ Rev. \ Lett. {\bf 93}, 205001

\bibitem{}Subramanian, K., Shukurov, A., Haugen, N.E.L.: 2005, astro-ph/0505144

\bibitem{}Vachaspati, T.: 2001, Phys. Rev. Lett. {\bf 87}, 251302

\bibitem{}Vall{\'e}e, J.P.: 2004, New Astron. Rev. {\bf 48}, 763

\bibitem{}Verma, M., Ayyer, A.: 2003, nlin.CD/0308005

\bibitem{}Vishniac, E., Cho, J.: 2001, Astrophys. J. {\bf 550}, 752

\bibitem{}Vishniac, E., Lazarian, A., Cho, J.: 2003,
Lect. Notes Phys. {\bf 614}, 376

\bibitem{}Widrow, L.: 2002, Rev. Mod. Phys. {\bf 74}, 775.

\end{thebibliography}
\end{document}